\documentclass[aps,prl,twocolumn]{revtex4}
\usepackage{graphicx}
\usepackage{indentfirst}
\usepackage{braket}
\usepackage{float}
\usepackage{CJK}
\usepackage{esint}
\usepackage{color}
\usepackage{subfigure}
\usepackage{amsfonts}

\begin{document}

\title{Dynamical topology and statistical properties of spatiotemporal chaos}

\author{Quntao Zhuang$^1$}
\author{Xun Gao$^1$}
\author{Qi Ouyang$^{1,2,3}$}
\author{Hongli Wang$^{1,2}$\footnote[1]{To whom correspondence should be addressed, E-mail:
hlwang@pku.edu.cn}}
\address{$^1$State key Laboratory for Mesoscopic Physics and School of Physics, Peking University, Beijing
100871; $^2$Center for Quantitative Biology, Peking University,
Beijing 100871; $^3$The Peking-Tsinghua Center for Life Sciences at
School of Physics, Beijing 100871, China}

\begin{abstract}
For spatiotemporal chaos described by partial differential
equations, there are generally locations where the dynamical
variable achieves its local extremum or where the time partial
derivative of the variable vanishes instantaneously. To a large
extent, the location and movement of these topologically special
points determine the qualitative structure of the disordered states.
We analyze numerically statistical properties of the topologically
special points in one-dimensional spatiotemporal chaos. The
probability distribution functions for the number of point, the
lifespan, and the distance covered during their lifetime are
obtained from numerical simulations. Mathematically, we establish a
probabilistic model to describe the dynamics of these topologically
special points. In despite of the different definitions in different
spatiotemporal chaos, the dynamics of these special points can be
described in a uniform approach.

\end{abstract}


\maketitle

\newpage

\textbf{Spatiotemporal chaos is often mediated by certain types of
dynamical spatial structures or topological features such as defects
in defect-mediated turbulence. The dynamics of such special entities
characterizes the disordered states and presents a simplified
description of spatiotemporal chaos. For the more general
spatiotemporal chaos that are not featured by defects and apparent
geometrical structures, it is a challenge to obtain a quantitative
mathematical characterization that records or preserves
qualitatively the geometric structures of the complex patterns. In
this paper, we propose to use the topologically special points such
as extremum and critical points in disordered states to characterize
qualitative properties of spatiotemporal chaos. To a large extent,
the location and movement of these topologically special points
determine the qualitative structure of the disordered states. We
calculate statistical properties of the extremum and critical points
with the Brusselator model and the complex Ginzburg-Landau equation.
A probabilistic model that can regenerate the simulation results is
proposed.}

\newpage

Spatially extended dynamical systems may exhibit spatiotemporal
chaos (STC) characterized by a finite correlation length in both
space and time \cite{STC}. While self-organization in pattern
formation systems has been intensively investigated in the last
decades, high-dimensional disordered states are still poorly
understood. The main challenge in this field has been to reveal the
onset of STC \cite{Ruelle,bif97,bif06} and to establish methods to
characterize such disordered states
\cite{STC,Manneville85,Egolf1995}. In many spatially extended
systems, spatiotemporal chaos is characterized by the presence of
defects. For such disordered states generally referred to as
defect-mediated turbulence (DMT), a statistical description of
defect dynamics has been used as a unifying approach to characterize
such STCs \cite{Gil90,bar,Daniels02,Daniels03,young,Kapral,hlwang04,
hlwang09,mikhailov10,mikhailov06,zhan08}. Numerical and theoretical
analyses have been carried out in the complex Ginzburg-Landau
equation \cite{Gil90,hlwang04}, the FitzHugh-Nagumo-type system
\cite{bar}, the Willamowski-Rossler model \cite{Kapral}, the model
for rotating non-Boussinesq convection \cite{young}, and catalytic
surface reactions \cite{mikhailov10}. Experimental studies of defect
statistics have been also reported from electro-convection in liquid
crystals \cite{Daniels02,Daniels03}, the catalytic surface reaction
\cite{mikhailov06}, and the Belousov-Zhabotinsky
reaction\cite{hlwang09}.

Similar to the defects in DMTs, other spatial structures or
topological features have been used to quantify the dynamical
complex patterns. Such examples have been reported from algebraic
topology for measuring the complexities in disordered states
\cite{Gameiro04}, instantaneous stagnation points with vanishing
velocity in the velocity field of fluids \cite{Fluid07}, and
self-replicating spots in the Gray-Scott model
\cite{Pearson93,hlwang07}.

In order to understand and quantify the more general STCs that are
not featured by defects and apparent geometrical structures, we here
propose a quantitative mathematical characterization that records or
preserves the dynamical structures of disordered states. For STCs
described by partial differential equations, there are generally
locations $r_0$ in the scalar field $X(r,t)$ where the dynamical
quantity achieves its local extremum, \emph{i.e.}, $\frac{\partial
X(r,t)}{\partial r}|_{r_0}=0$, as depicted in Fig. 1a. These
instantaneous maximum or minimum points preserve the geometrical
structures of the disordered states (Fig. 1c). If the locations and
movements of all of these special points are known, the pattern can
be primarily determined. Similarly, as in the velocity field of
fluid flows \cite{Fluid07}, the spatiotemporal chaotic field
$X(r,t)$ can be characterized by locations $r_0$ where the time
partial derivative of the dynamical quantity vanishes
instantaneously, \emph{i.e.}, $\frac{\partial X(r,t)}{\partial
t}|_{r_0}=0$, as demonstrated in Fig. 1b. These critical points
carry the bulk of the information contained in the disordered
patterns (Fig. 1d). Such defined topologically special points, which
are present even in ordered states, are distinct from the
topological defects in previously studied DMTs. They are created and
annihilated irregularly in pairs, changing simultaneously the
topological structures of the disordered states.  To a large extent,
the location and movement of these points determine the qualitative
structure of the disordered states.

\begin{figure}[tbp]
    \centering
    \includegraphics[width=8.5cm]{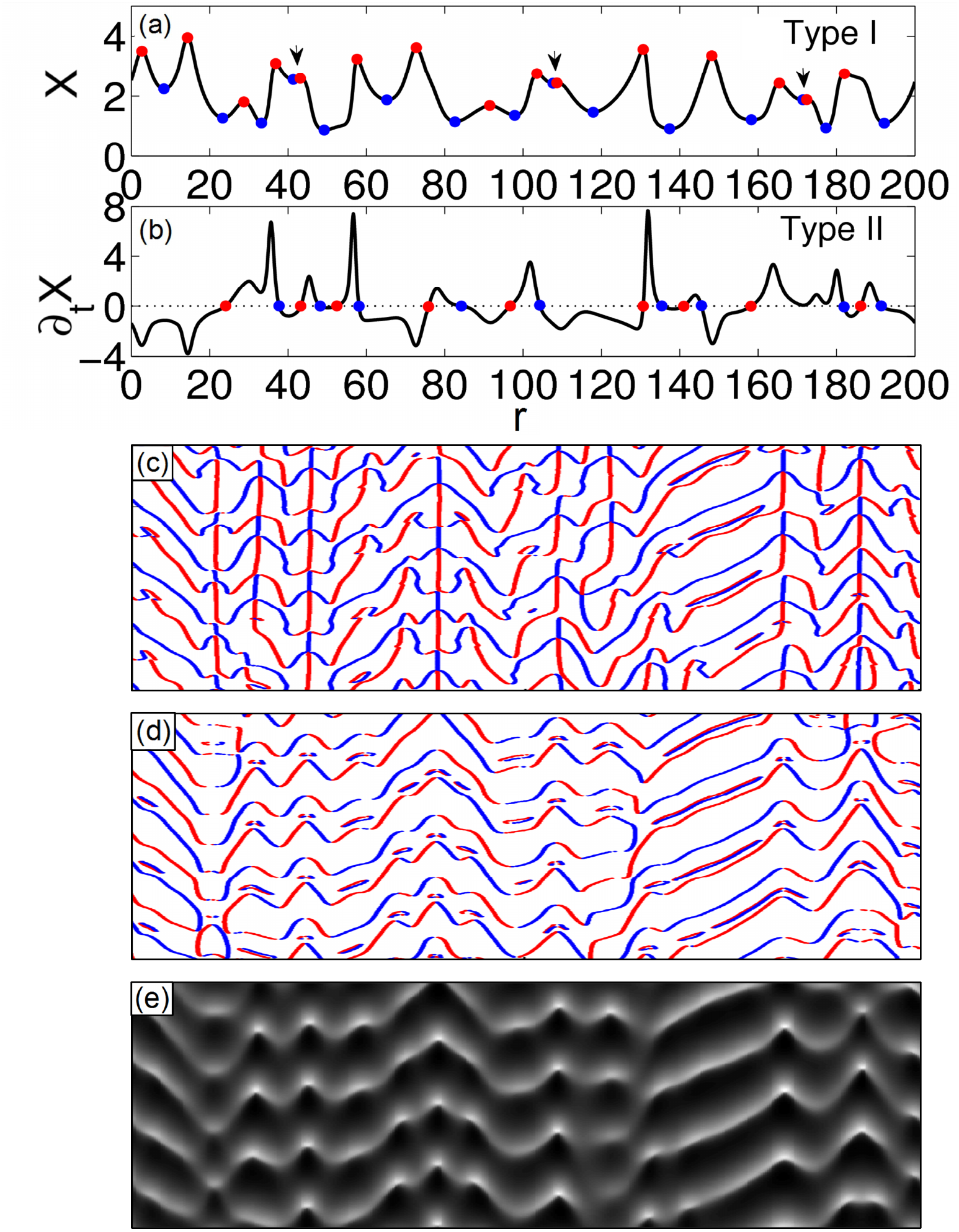}
    \caption{One-dimensional spatiotemporal chaos generated from the reaction-diffusion
system with Brusselator reaction kinetics (Eqs. 1 and 2). (a)
Snapshot of $X(r,t)$ and local maximum (red) and local minimum
(blue) points; (b) Snapshot of $\partial_t X(r,t)$ and critical
points where $\partial_t X(r,t)=0$, with red for
$\partial_r[\partial_t X(r,t)]>0$ and blue for
$\partial_r[\partial_t X(r,t)]<0$. (c) and (d) are space-time plots
(horizontal space and vertical time) depicting the time evolutions
of the topological special points in (a) and (b), respectively. (e)
The space-time map of $X(r,t)$ coded in gray scales. The arrows in
(a) indicate the red-blue pairs being created or annihilated in the
disordered states. Parameters: $A=2.0, B=5.5, D_X=1, D_Y=0$.
Periodic boundary conditions are used in simulations.
    }
\end{figure}

We analyze numerically statistical properties of the topologically
special points in one-dimensional STCs with the Brusselator model
and the complex Ginzburg-Landau equation (CGLE). The probability
distribution functions (PDF) for the number of extremum points and
that of critical points are obtained from numerical simulations.
PDFs for the lifespan as well as the distance covered during the
lifetime are also calculated. Mathematically, we establish a
probabilistic model to describe the dynamics of these topologically
special points. The analysis gives results that are in good
agreement with numerical simulations. In despite of the different
definitions in different STCs, the dynamics of these special points
can be described uniformly.

\begin{figure}[tpb]
    \centering
    \includegraphics[width=7cm]{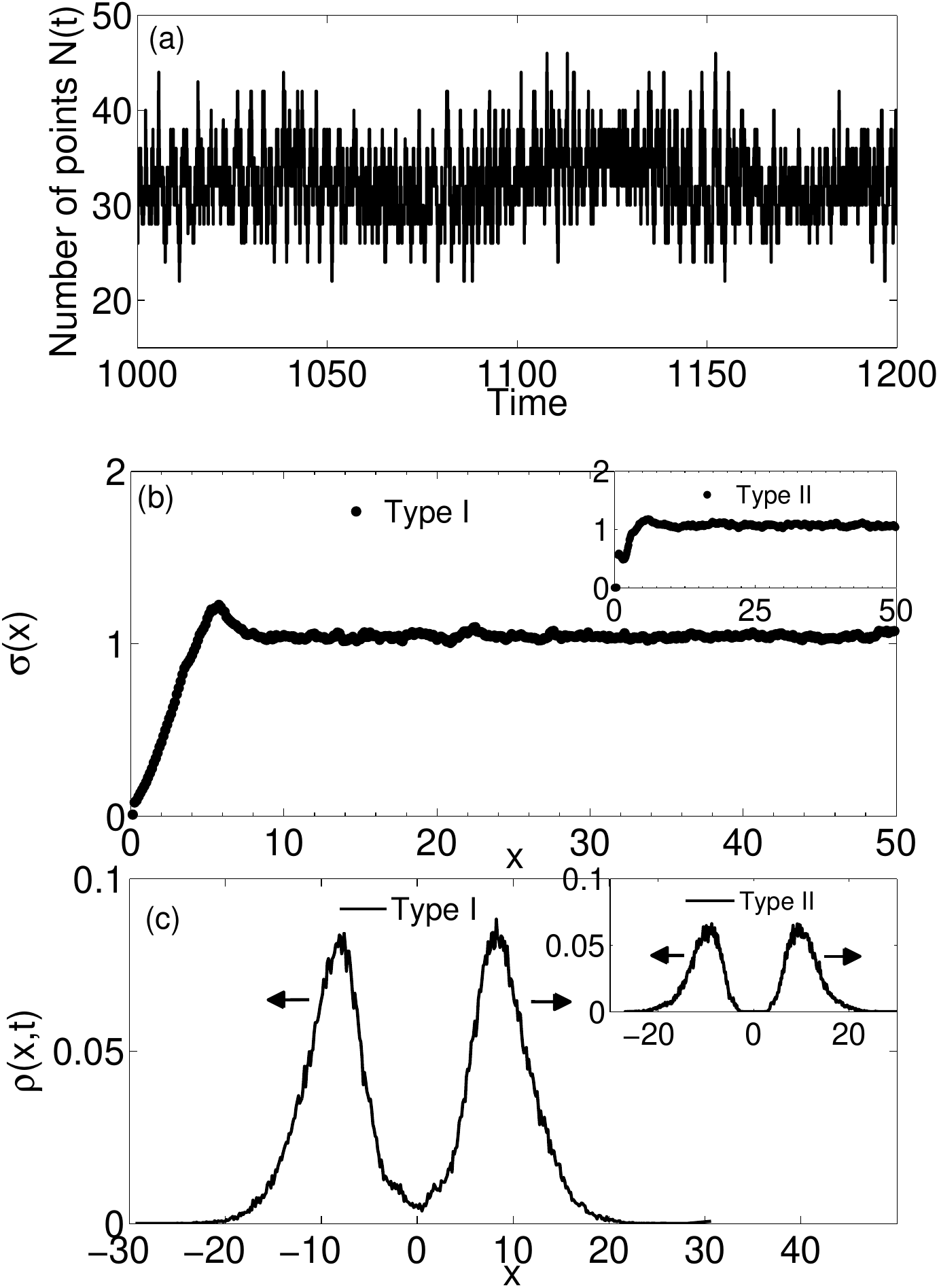}
    \caption{(a) Time series $N(t)$ for the number of extremum points.
    (b) The normalized pair correlation function
$\sigma(r)$ for the extremum points (inset for the critical points).
(c) Bimodal probability density function $\rho(x,t=2.5)$ for
extremum points (inset for critical points). The arrows indicate the
directions $\rho(x,t)$ is splitting. All the simulation results are
obtained from the Brusselator model with the same parameters in Fig.
1.}
\end{figure}

\emph{Models and numerical results}.--- The reaction-diffusion
system with Brusselator reaction kinetics is a standard model system
for the investigation of self-organized structures in nonlinear
chemical reactions, which is governed by the following set of
partial differential equations \cite{Brusselator},
\begin{eqnarray}
\partial_tX&=&D_X\partial_r^2X+X^2Y-(B+1)X+A,\label{qer} \\
\partial_tY&=&D_Y\partial_r^2Y-X^2Y+BX.
\end{eqnarray}
Figure 1 demonstrates the chemical turbulence generated by Eqs. 1
and 2 with parameters $A=2.0, B=5.5, D_X=1, D_Y=0$. Figure 1a and 1b
depict the snapshots of the one-dimensional fields $X(r,t)$ and
$\frac{\partial X(r,t)}{\partial t}$, respectively. The maximum
(minimum) points in Fig. 1a are denoted as red (blue). The dots in
Fig. 1b are for the critical points where $\frac{\partial
X(r,t)}{\partial t}=0$, with the red or blue for locations where
$\partial_r[\partial_t X(r,t)]>0$ or $\partial_r[\partial_t
X(r,t)]<0$. As these special points demonstrated in Fig. 1a and 1b
are always created and annihilated in red-blue pairs, the red and
the blue can be denoted with positive (+) and negative (-) signs
similar to the positive and negative charges for topological defects
in DMTs. The space-time plots in Fig. 1c and 1d demonstrate the time
evolutions of the topologically special points in Fig. 1a and Fig.
1b, respectively. The gray-scaled space-time pattern in Fig. 1e is
generated from $X(r,t)$. To a large extent, the complexities of the
disordered states (Fig. 1e) are primarily preserved or recorded in
the time evolutions of these topological points (Fig. 1c and 1d).
The definitions of the special points for the Brusselator can be
easily extended to the CGLE and other systems. The CGLE which is one
of the most-studied nonlinear equations in physics is given by
\cite{GL},
\begin{eqnarray}
\partial_tA&=&A+(1+ic_1)\partial_r^2A-(1-ic_3)|A|^2A.
\end{eqnarray}
Beyond the Benjamin-Feir instability, three types of chaotic
behavior are possible in the one-dimensional CGLE, namely phase
turbulence, space-time defect turbulence \cite{cgle_chaos_1_2}, and
intermittency \cite{cgle_chaos3}. As the emphasis of this paper is
on the presentation of the technique, we here concentrate on the
topological special points as defined for the Brusselator without
discrimination in the types of STCs.

\begin{figure}[t]
    \centering
    \includegraphics[width=8cm]{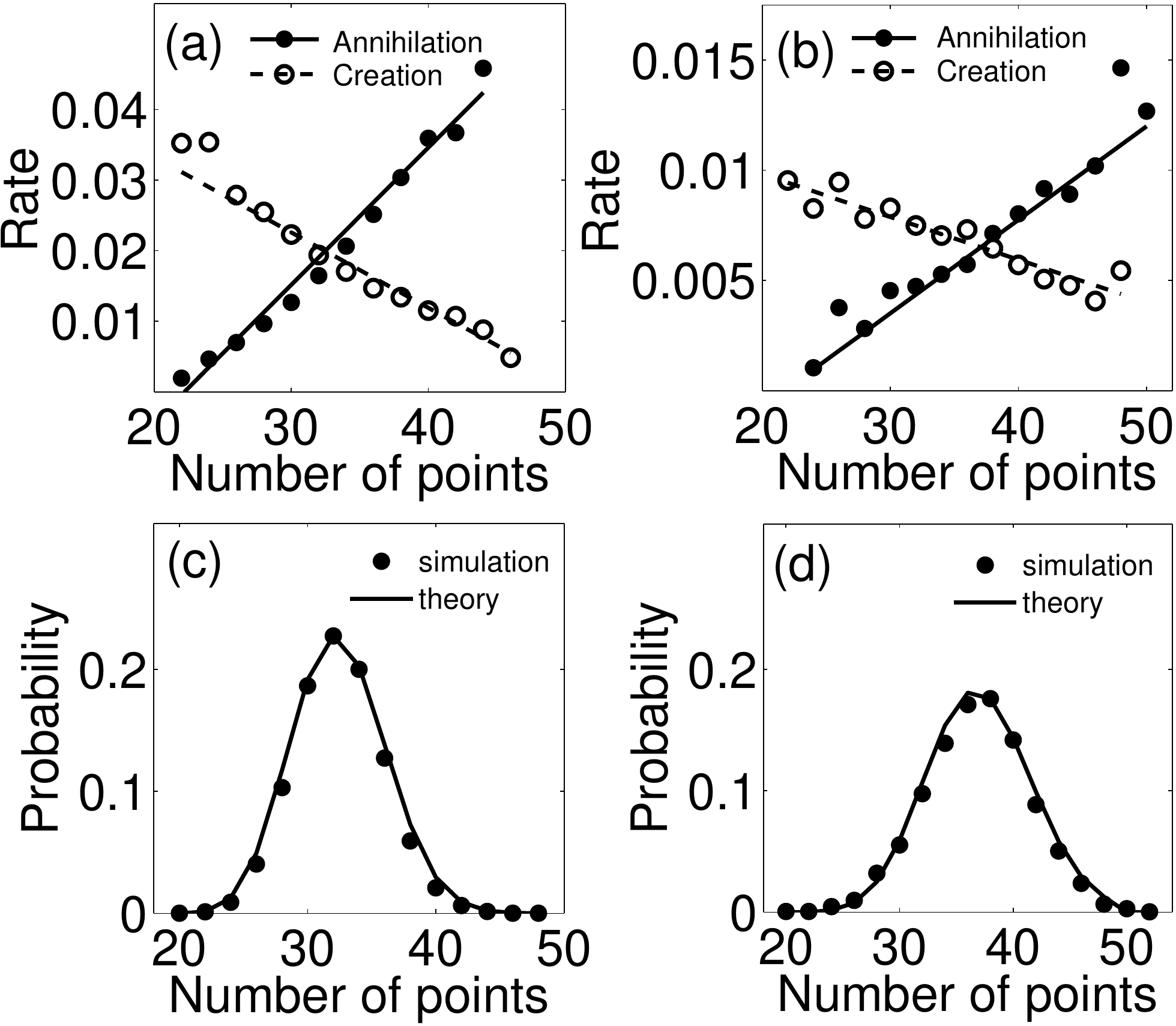}
    \caption{Numerically simulated gain and loss rates, which are approximately linear functions,
    for extremum points obtained from the Brusselator model (a)
     and critical points from the CGLE (b), and probability distribution functions for the
     number of extremum points obtained from the Brusselator (c) and of critical points from the CGLE.
     Theoretical PDFs from Eq. 4 are depicted as solid lines.
    }
\end{figure}

\begin{figure*}[t]
    \centering
    \includegraphics[width=16cm]{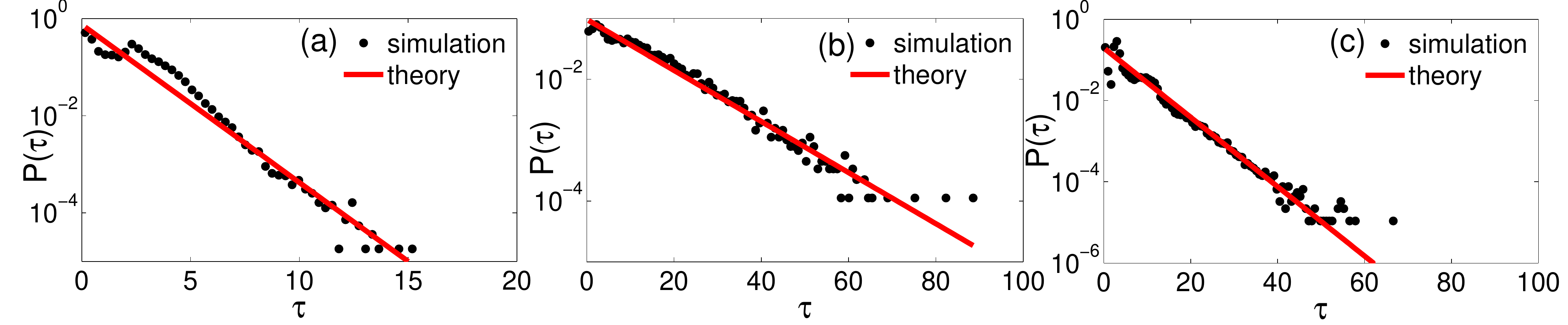}
    \caption{Lifetime distribution $P(\tau)$ of the topologically special
    points obtained from numerical
simulations (black dots) and from theoretical prediction of Eq.
9(red line). (a) Extremum points in the Brusselator model,
$C=0.7482$; (b) Critical points in the CGLE, $C=0.1964$. (c)
Critical points in the autocatalator model, $C=0.09669$. }
\end{figure*}

As a simplified representation of spatiotemporal chaos, the
dynamical behavior of extremum and critical points are more
convenient to describe and cope with. The points perform diffusive
and drift motions in the space after they are created in pair, are
destroyed when two neighboring points with opposite sign bump into
each other. The processes occur continuously and indefinitely, and
the spatiotemporal pattern looks like a chaotic ``soup". The number
of extremum or critical points in the disordered pattern undergoes
fast fluctuations. Figure 2a illustrates a time series $N(t)$ of the
number of extremum point calculated from the Brusselator. The fast
stochastic fluctuations in $N(t)$ is a manifestation of the disorder
and incoherence in the turbulence. Figure 2b demonstrates the pair
correlation functions for the extremum and critical points, showing
short range interactions between these points.

The diffusive and drift motion of extremum and critical points can
be examined from the time dependent probability density function
$\rho(x,t)$ for the distance $x$ that a point travels from the
location where it is created at the time $t$ from the moment it is
created. This can be calculated by tracking the points from their
creation to annihilation processes. Apparently the initial
distribution $\rho(x,0)$ is a delta function $\delta(x)$. As
demonstrated in Fig. 2c, it will then split as the pairs of point
that are newly born with opposite signs move apart, forming a
bimodal distribution $\rho(x,t)$. By monitoring the creation and
annihilation events, the gain and loss rates can be obtained
numerically (Fig. 3a and 3b). Both are approximately linear
functions in the number of points in the system. From the time
series $N(t)$, the probability density function (PDF) $P(N)$ which
is the probability for finding $N$ points in the disordered states
can be derived. Figure 3c and 3d (dots) demonstrate the PDFs for the
extremum and critical points obtained from the Brusselator and the
CGLE, respectively.

By monitoring the creation, movement, and annihilation processes of
these points, we derive two extra properties: the probability
density function $P(\tau)$ for the life span $\tau$ of the special
points, and the distribution $P(\lambda)$ of the displacement
$\lambda$ (i.e., the distance a point covers during its lifetime).
Figure 4 (dots) demonstrates the life span distributions for
extremum and critical points calculated separately from the
Brusselator model (Fig. 4a) and the CGLE (Fig. 4b). They are
uniformly exponential decays. Figure 5 (dots) depicts the
distribution of displacement $\lambda$ for extremum and critical
points in the Brusselator model (Fig. 5a) and CGLE (Fig. 5b),
respectively. The distributions are also uniformly tent-shaped. The
left-right symmetry in the distributions is due to the symmetrical
splitting movement of the special points after they are created.

\emph{Theoretical analysis}.--- We now seek to establish a
probabilistic model to generate quantitatively the numerical
results. The topologically special points we defined here are
similar to defects in DMTs. The master equation description of
defect dynamics \cite{Gil90} can be extended in straightforward to
the situation we consider here. The probability distribution
function for the number of extremum or critical point can be readily
expressed as \cite{Gil90},
\begin{equation}
P(N)=P(0)\prod_{j=0}^{j=N-1}\frac{\Sigma_+(j)}{\Sigma_-(j+1)},
\end{equation}
where $\Sigma_+$ and $\Sigma_-$ are, respectively, gain and loss
rates that define the transition rates of the Markov chain. From our
simulations, $\Sigma_+$ and $\Sigma_-$ for the Brusselator and the
CGLE are well approximated by linear dependence on $N$,
$\Sigma_+(N)=c_0+c_1N$, $\Sigma_-(N)=a_0+a_1N$ (solid lines in Fig.
3a and 3b). The resulting distributions from Eq. 4 are depicted in
Fig. 3c and 3d (curves) which are in good agreement with simulations
in despite that the special points are different and in different
models.

\begin{figure*}[t]
    \centering
    \includegraphics[width=16cm]{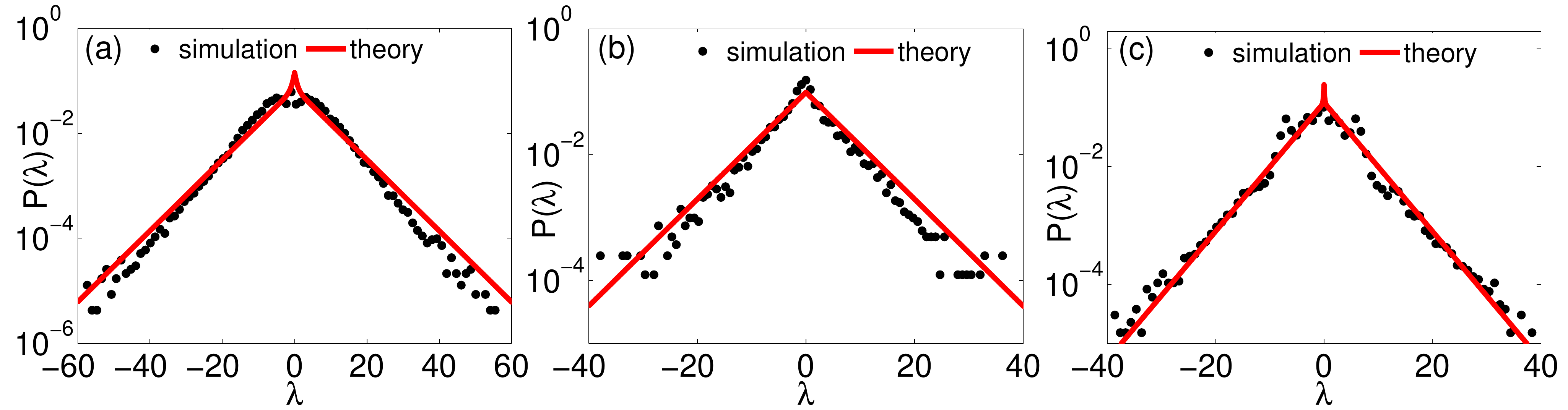}
    \caption{Displacement distribution $P(\lambda)$ for the special points
     obtained from
numerical simulations (black dots) and from theoretical predictions
of Eq. 11 (red line). (a) for extremum points in the Brusselator
model, $v=4.331, D=2.9925$; (b) for critical points in the CGLE,
$v=0.764$, $D=0.0601$; (c) for critical points in the autocatalator
model, $v=3.0$, $D=2.526$. The values of $C$ are as in Fig. 4.}
\end{figure*}

Our finding that the creation and annihilation rate of the special
points are both linear with $N$ is distinct from previous findings
in 2D DMT where the annihilation rate is proportional to $N^2$
\cite{Gil90}. The difference might be due to that different
dimensions have been considered. The $N^2$ annihilation rate is
intuitive in 2D systems. In 1D systems as we consider, the positive
spots and negative spots are alternatively distributed in the 1D
space. As an elementary annihilation or creation event occurs
between two spots with opposite signs, it is correlated with two
neighboring spots. The annihilation/creation rates should thus be
proportional to the number of spot pairs in the space and would
result in the linear dependence on $N$ as we observed.

In order to analyze the behaviors of topologically special point in
more detail, an equation for the probability density function
$\rho(x,t)$ needs to be established. As observed in our simulations,
the extremum and critical points perform drifting and diffusive
motion (Fig. 2c), and suffer random extinctions during these
processes. To describe these properties, we propose a modified
Fokker-Planck equation for the time evolution of $\rho_+(x,t)$ and
$\rho_{-}(x,t)$ with $\rho(x,t)=(\rho_+(x,t)+\rho_-(x,t))/2$,
\begin{equation}
\partial_t\rho_{\pm}(x,t)=\{\mp v \partial_x
+D \partial_x^2 -[c\sum_i \delta(x-x_i(t))]\}\rho_{\pm}(x,t).
\end{equation}
The sign $\pm$ describes the symmetrical splitting of $\rho(x,t)$
into $\rho_+$ and $\rho_-$. In Eq. 5, constant drifting and
diffusion are assumed. The right most term is used to model
interaction decays, where $x_i(t)$ is the location of a special
point leading to the decay. In a mean-field approximation, the last
term is proportional to the normalized pair correlation function
$\sigma(x)$ (Fig. 2b). One has $<c\sum_i
\delta(x-x_i(t))>_t=c\sigma(x)$. Due to that $\sigma(x)$ approaches
a constant value beyond a small spatial correlation in STC, one has,
\begin{equation}
\frac{\partial }{\partial t}\rho_{\pm}(x,t)=\{\mp
v\frac{\partial}{\partial x}+D\frac{\partial^2 }{\partial
x^2}-C\}\rho_{\pm}(x,t).
\end{equation}
The above equation has the solution
\begin{equation}
\rho_{\pm}(x,t)=\frac{1}{\sqrt{4\pi Dt}}\exp[-\frac{(x\mp
vt)^2}{4Dt}]\exp(-Ct).
\end{equation}
The probability $\rho(x,t)$ is thus,
\begin{equation}
\rho(x,t)=(\rho_+(x,t)+\rho_-(x,t))/2.
\end{equation}
The life span distribution $P(\tau)$ can be calculated directly from
the decreasing of the total probability,
\begin{equation}
P(\tau)=-\frac{d}{d\tau}\int_{-\infty}^\infty
\rho(x,\tau)dx=C\exp(-C\tau).
\end{equation}
The probability is exactly an exponentially decaying function where
$C$ depends on the detailed short-range interactions. This
theoretical result is in good agreement with numerical results as
shown in Fig. 4a and 4b.

To calculate the distribution of displacement $P(\lambda)$, we have
the probability density for the annihilation at location $x$ with
life time $\tau$ as $\rho_F(x,\tau)P(\tau)$, where
$\rho_F(x,t)=(\rho_+(x,t)+\rho_-(x,t))/2|_{C=0}$ is the solution of
Eq. 6 with $C=0$. The free path length distribution $P(\lambda)$ is,
\begin{equation}
P(\lambda)=\int_0^{\infty}d\tau\rho_F(\lambda,\tau)P(\tau).
\end{equation}
The integration leads to,
\begin{equation}
\begin{array}{l}
P(\lambda)=\frac{C}{\sqrt{4CD+v^2}}\exp[-\frac{\lambda}{2D}(\sqrt{4CD+v^
2})\frac{\lambda}{|\lambda|}]\\
\times(\exp[-v\lambda/2D]+\exp[v\lambda/2D])/2.
\end{array}
\end{equation}
For a specific disordered state, the drift velocity $v$ and
diffusion coefficient $D$ in the above equation can be determined
from the simulations of $\rho(x,t)$ (refer to Fig. 2c) using Eq. 8,
and the constant $C$ can be determined from the simulated life span
distribution according to Eq. 9. With these parameters, the
theoretical result predicted by Eq. 11 for the extremum and critical
points the Brusselator and in the CGLE demonstrated in Fig. 5a and
5b (curves), respectively, which also agree uniformly well with the
results obtained from direct simulations.

\emph{Summary and Conclusion.}--- In summary, we have proposed using
topologically special points such as extremum and critical points in
disordered states to characterize the qualitative properties of
spatiotemporal chaos that is not mediated by defects. The special
points can be denoted as positive or negative, and undergo pairwise
interactions, changing the topological structures of the disordered
states when they are created and annihilated. The dynamics of such
points are similar to the $A+B\rightleftharpoons C$ type chemical
reaction in 1D space \cite{Lythe}, where different reactants $A$ and
$B$ combine into resultant $C$ when they encounter and $C$
decomposes into $A$ and $B$. We have checked the dependence of the
average on the length of the system (data not presented here). The
average number of special points is proportional to the size of the
system and is therefore extensive in spatiotemporal chaos.

We demonstrated with the Brusselator model and the CGLE that the
dynamics of special points determining the qualitative properties of
STCs can be described in a probabilistic model, in spite of
different definitions of the special points and different dynamical
systems. Specifically, the previous statistical description of
defect dynamics in defect-mediated turbulence was extended to the
general spatiotemporal chaos. A modified Fokker-Planck equation was
proposed as a uniform description for the drifting, diffusive and
extinctions processes. Theoretical analysis gives explicitly
expressions for the lifespan distribution and the displacement
distribution, and predicts correctly numerical simulation results.

The definition of the topological and dynamical special points and
the probabilistic approach that we proposed for 1D STCs in the
Brusselator and CGLE can be extended to other disordered states. For
instance, we have analyzed the chemical turbulence which emerges
from the instability of a traveling wave. The model is a
two-variable, cubic autocatalator with equal diffusivities
 of the species which is described by the following equations
\cite{autocatalator},
\begin{eqnarray}
\partial_t\alpha&=&\delta\partial_z^2\alpha+1-\alpha-\mu\alpha\beta^
2,\\
\partial_t\beta&=&\partial_z^2\beta+\mu\alpha\beta^2-\phi\beta.
\end{eqnarray}
As demonstrated in Fig. 4c, the lifetime distribution of the
critical points obtained from numerical simulations (black dots) in
the STC regime, with $\delta=1,\phi=3.0, \mu=38.0$, is in
satisfactory agreement with that predicted by Eq. 9 (red line).
Figure 5c depicts the same agreement between the numerical
simulation and theoretical prediction of the displacement
distribution.

In 2D defect-mediated turbulence, the defect is a convenient
characterization of turbulence as previously investigated. The
method present here is suitable only for 1D systems. It is
convenient to define and handle topologically special points in 1D
systems. In 2D systems, the special points as we defined in 1D would
be no longer isolated points, but could be connected into 1D lines
or filaments.

The work is financially supported by the NSFC (10721403, 11074009,
10774008, 11174013) and MSTC (2009CB918500).


\begin{references}

\bibitem{STC} M. C. Cross and P. C. Hohenberg, Rev. Mod. Phys. 65,
851 (1993).

\bibitem{Ruelle} D. Ruelle and F. Takens, Commun. Math. Phys. \textbf{20},
167 (1971).

\bibitem{bif97} Z. Qu, J. N. Weiss, and A. Garfinkel, Phys. Rev. Lett.
\textbf{78}, 1387 (1997).

\bibitem{bif06} F. Brochard, E, Gravier, and G. Bonhomme, Phys. Rev. E \textbf{73},
036403 (2006).

\bibitem{Manneville85} P. Manneville, in\emph{ Macroscopic Modelling of
Turbulent Flows}, Springer, Berlin, 1985.

\bibitem{Egolf1995} D. A. Egolf and H. S. Greenside, Nature, \textbf{369},
129 (1994); \emph{ibid}, Phy. Rev. Lett. \textbf{74}, 1751 (1995).

\bibitem{Gil90} L. Gil, J. Lega, and J.L. Meunier, Phys. Rev. A
{\bf 41}, 1138 (1990).

\bibitem{hlwang04} H. Wang, Phys. Rev. Lett. \textbf{93}, 154101
(2004);

\bibitem{bar} M. Hildebrand, M. B\"{a}r, and M. Eiswirth, Phys. Rev.
Lett. \textbf{75}, 1503 (1995).

\bibitem{Kapral} J. Davidsen and R. Kapral, Phys. Rev. Lett. {\bf 91},
058303 (2003);

\bibitem{young} Y. N. Young and H. Reicke, Phys. Rev. Lett. \textbf{90}, 134502
(2003).

\bibitem{mikhailov10} D. Krefting and C. Beta, Phys. Rev. E
\textbf{81}, 036209 (2010).

\bibitem{Daniels02} K. E. Daniels and E. Bodenschatz, Phys. Rev.
Lett. {\bf 88}, 034501 (2002).

\bibitem{Daniels03} K. E. Daniels and E. Bodenschatz, Chaos {\bf 13}, 55
(2003).

\bibitem{mikhailov06} C. Beta, A. S. Mikhailov, H. H. Rotermund, and G. Ertl, Europhys.
Lett. \textbf{75}, 868 (2006).

\bibitem{hlwang09} C. Qiao, H. Wang, and Q. Ouyang, Phys. Rev. E \textbf{79}, 016212
(2009).

\bibitem{zhan08} J. Davidsen, M. Zhan, and R. Kapral, Phys. Rev. Lett. \textbf{101},
208302 (2008).

\bibitem{Gameiro04} M. Gameiro, K. Mischaikow£¬and W. Kalies, Phys. Rev. E
\textbf{70}, 035203(R) (2004).

\bibitem{Fluid07} N. T. Ouellette and J. P. Gollub, Phys. Rev. Lett \textbf{99},
194502 (2007); \emph{ibid}, Phys. Fluids, \textbf{20}, 064104
(2008).

\bibitem{Pearson93} J. E. Pearson, Science \textbf{261}, 189 (1993).

\bibitem{hlwang07} H. Wang and Q. Ouyang, Phys. Rev. Lett. \textbf{99}, 214102 (2007).

\bibitem{Brusselator} S. Chakravarti, M. Marek, and W. H. Ray, Phys. Rev. E \textbf{52},
2407 (1995).

\bibitem{GL} I. Aranson, L. Kramer, Rev. Mod. Phys. \textbf{74}, 99 (2002).

\bibitem{cgle_chaos_1_2} B. I. Shraimana, A. Pumir, W. van Saarloos, P. C.
Hohenberg, H. Chate, M. Holen, Physica D \textbf{57}, 241 (1992).

\bibitem{cgle_chaos3} H Chate, Nonlinearity \textbf{7}, 185 (1994).

\bibitem{Lythe} S. Habib, G. Lythe, Phys. Rev. Lett. \textbf{84}, 1070
(2000); G. Lythe, Physica D \textbf{222}, 159 (2006).

\bibitem{autocatalator} J. H. Merkin, V. Petrov, S. K. Scott, K. Showalter, Phys. Rev. Lett.
\textbf{76}, 3 (1996).

\end{references}
\end{document}